\documentclass[conference]{IEEEtran}

\usepackage{amsthm}
\usepackage[fleqn]{amsmath}

\ifCLASSINFOpdf
   \usepackage[pdftex]{graphicx}
\else
   \usepackage[dvips]{graphicx}
   \graphicspath{{../eps/}}
\fi
\begin{document}
\title{Secure and Reliable Transmission with Cooperative Relays in Two-Hop Wireless Networks}

%\author{\IEEEauthorblockN{Michael Shell}
%\IEEEauthorblockA{School of Electrical and\\Computer Engineering\\
%Georgia Institute of Technology\\
%Atlanta, Georgia 30332--0250\\
%Email: http://www.michaelshell.org/contact.html} \and
%\IEEEauthorblockN{Homer Simpson}
%\IEEEauthorblockA{Twentieth Century Fox\\
%Email: homer@thesimpsons.com} \and \IEEEauthorblockN{James Kirk\\
%and Montgomery Scott}
%\IEEEauthorblockA{Starfleet Academy\\
%San Francisco, California 96678-2391\\


\author{\IEEEauthorblockN{Yulong Shen\IEEEauthorrefmark{1}\IEEEauthorrefmark{4},
Xiaohong Jiang\IEEEauthorrefmark{2}, Jianfeng
Ma\IEEEauthorrefmark{1} and Weisong Shi\IEEEauthorrefmark{3}}
\IEEEauthorblockA{\IEEEauthorrefmark{1}School of Computer Science
and Technology, Xidian University, China}
\IEEEauthorblockA{\IEEEauthorrefmark{2}School of Systems Information
Science, Future University Hakodate, Japan}
\IEEEauthorblockA{\IEEEauthorrefmark{3}Department of Computer
Science, Wayne State University, USA}
\IEEEauthorblockA{\IEEEauthorrefmark{4}Email:ylshen@mail.xidian.edu.cn
} }

\maketitle

\begin{abstract}
%\boldmath
This work considers the secure and reliable information transmission
in two-hop relay wireless networks without the information of both
eavesdropper channels and locations. While the previous work on this
problem mainly studied infinite networks and their asymptotic
behavior and scaling law results, this papers focuses on a more
practical network with finite number of system nodes and explores
the corresponding exact results on the number of eavesdroppers the
network can tolerant to ensure a desired secrecy and reliability.
For achieving secure and reliable information transmission in a
finite network, two transmission protocols are considered in this
paper, one adopts an optimal but complex relay selection process
with less load balance capacity while the other adopts a random but
simple relay selection process with good load balance capacity.
Theoretical analysis is further provided to determine the exact and
maximum number of independent and also uniformly distributed
eavesdroppers one network can tolerate to satisfy a specified
requirement in terms of the maximum secrecy outage probability and
maximum transmission outage probability allowed.

%show the the maximum number of eavesdroppers that can be tolerated to achieve desired secrecy in terms of transmission outage and secrecy outage probability.and explore the corresponding exact  (rather than order sense) results .studies the problem of information transmission secrecy in two-hop wireless networks with limited cooperative relays. Two transmission protocols with optimal relay selection in \cite{IEEEhowto:Goeckel} and random relay selection in this paper are considered to ensure the information transmission secrecy in cooperative relay wireless networks. This paper provides a theoretic analysis for these two protocols on the number of eavesdroppers that can be tolerated to achieve desired secrecy in terms of transmission outage and secrecy outage probability. Moreover, we present the more effectively information transmission with the joint use of these two protocols according the application scenario while achieving the reliability and secrecy of transmission, since the protocol in \cite{IEEEhowto:Goeckel} is suitable for the small size and sufficient energy wireless networks, while the protocol in this paper is more suitable for the large size and limited energy wireless networks.

\end{abstract}

%\begin{IEEEkeywords}
%Wireless Networks, Cooperative Relay, Information-Theoretic
%Security.
%\end{IEEEkeywords}

% \fi
\IEEEpeerreviewmaketitle

\section{Introduction}

Two-hop ad hoc wireless networks, where each packet travels at most
two hops (source-relay-destination) to reach its destination, has
been a class of basic and important networking scenarios
\cite{IEEEhowto:Sathya}. Actually, the analysis of basic two-hop
relay networks serves as the foundation for performance study of
general multi-hop networks. Due to the promising applications of ad
hoc wireless networks in many important scenarios (like battlefield
networks, emergency networks, disaster recovery networks), the
consideration of secrecy (and also reliability) in such networks is
of great importance for ensuring the high confidentiality
requirements of these applications. This paper focuses on the issue
of secure and reliable information transmission in the basic two-hop
ad hoc wireless networks.

Traditionally, the information security is provided by adopting the
cryptography approach, where a plain message is encrypted through a
cryptographic algorithm that is hard to break (decrypt) in practice
by any adversary without the key. While the cryptography is
acceptable for general applications with standard security
requirement, it may not be sufficient for applications with a
requirement of strong form of security (like military networks and
emergency networks). This is because that the cryptographic approach
can hardly achieve everlasting secrecy, since the adversary can
record the transmitted messages and try any way to break them
\cite{IEEEhowto:Talbot}. That is why there is an increasing interest
in applying signaling scheme in physical layer to provide a strong
form of security, where a degraded signal at an eavesdropper is
always ensured such that the original data can be hardly recovered
regardless of how the signal is processed at the eavesdropper. We
consider applying physical layer method to guarantee secure and
reliable information transmission in the two-hop wireless networks.

By now, a lot of research efforts have been dedicated to providing
security through physical layer methods. A power control scheme is
proposed in \cite{IEEEhowto:Morr} to ensure that an eavesdropper can
never reach its desired signal-to-noise-plus-interference ratio
(SINR). However, such scheme is not effective when the eavesdropper
has a better channel than the receiver. The technique of artificial
noise generation has also been widely explored to jam the
eavesdroppers and provide secure transmission in the relay
communications
\cite{IEEEhowto:Goel}\cite{IEEEhowto:Lai}\cite{IEEEhowto:Yuksel}\cite{IEEEhowto:Negi}.
Recently, the cooperative jamming through node cooperation has been
demonstrated to be efficient in ensuring physical layer security
\cite{IEEEhowto:Vasudevan}\cite{IEEEhowto:He}\cite{IEEEhowto:Dong}.
It is notable that these schemes generally reply on the knowledge of
eavesdropper channels and locations to jam eavesdroppers. In
practice, however, it is difficult to gain such information,
specifically in untrusted network environment. To address this
constraint, a cooperative protocol based on artificial noise
generation and multi-user diversity has been proposed recently in
\cite{IEEEhowto:Goeckel} to achieve secure transmission in two-hop
wireless networks without the knowledge of eavesdropper channels and
locations. In particular, the asymptotic behavior of such
cooperative protocol in a network has been reported there to
illustrate how the number of eavesdroppers the network can tolerate
scales as the number of system nodes there tends to infinite.

This paper focuses on applying the relay cooperation scheme to
achieve secure and reliable information transmission in a more
practical finite two-hop wireless network without the knowledge of
both eavesdropper channels and locations. The main contributions of
this paper as follows:

1) For achieving secure and reliable information transmission in a
more practical two-hop wireless network with finite number of system
nodes, we consider the application of the cooperative protocol
proposed in \cite{IEEEhowto:Goeckel} with an optimal and complex
relay selection process but less load balance capacity, and also
propose to use a new cooperative protocol with a simple and random
relay selection process but good load balance capacity.

2) Rather than exploring the asymptotic behavior and scaling law
results, this paper provides theoretic analysis of above both
cooperative protocols to determine the corresponding exact results
on the number of independent and also uniformly distributed
eavesdroppers one network can tolerate to satisfy a specified
requirement in terms of the maximum secrecy outage probability and
maximum transmission outage probability allowed.

The remainder of the paper is organized as follows. Section II
introduces the system models and two cooperative transmission
protocols considered in this paper. Section III provides theoretical
analysis and also related discussions of the two protocols, and
Section IV concludes this paper.

\section{System Models and Transmission Protocols}

\subsection{Network Model}
As illustrated in Fig.1 that we consider a network scenario where a
source node $S$ wishes to communicate securely with its destination
node $D$ with the help of multiple relay nodes $R_1$, $R_2$,
$\cdots$, $R_n$. In addition to these normal system nodes, there are
also $m$ eavesdroppers $E_1$, $E_2$, $\cdots$, $E_m$ that are
independent and also uniformly distributed in the network. Our goal
here is to ensure the secure and reliable information transmission
from source $S$ to destination $D$ under the condition that no real
time information is available about both eavesdropper channels and
locations.

\begin{figure}[!t]
\centering
\includegraphics[width=2in]{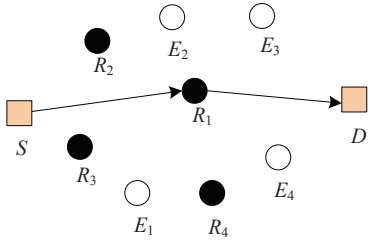}
\DeclareGraphicsExtensions. \caption{System scenario: Source $S$
wishes to communicate securely with destination $D$ with the
assistance of finite relays $R_1$, $R_2$, $\cdots$, $R_{n}$ ($n$=4
in the figure) in the presence of passive eavesdroppers $E_1$,
$E_2$, $\cdots$, $E_{m}$ ($m$=4 in the figure). Cooperative relay
scheme is used in the two-hop transmission. A assistant node is
selected randomly as relay ($R_1$ in the figure).} \label{System
scenario}
\end{figure}

\subsection{Transmission Model}

Consider the transmission from a transmitter $A$ to a receiver $B$,
and denote by $x_i^{\left(A\right)}$ the $i^{th}$ symbol transmitted
by $A$ and denote by $y_i^{\left(B\right)}$ the $i^{th}$ signal
received by $B$. We assume that all nodes transmit with the same
power $E_s$, path loss between all pairs of nodes is equal and
independent, and the frequency-nonselective multi-path fading from
$A$ to $B$ is a complex zero-mean Gaussian random variable. Under
the condition that all nodes in a group of nodes, $\mathcal {R}$,
are generating noises, the $i^{th}$ signal received at node $B$ from
node $A$ is determined as:

$$y_i^{\left(B\right)}=h_{A,B} \sqrt{E_s}x_i^{\left(A\right)} +
\sum_{A_i \in \mathcal {R}}
h_{A_i,B}\sqrt{E_s}x_i^{\left(A_i\right)} + n_i^{\left(B\right)},$$

where the noise $\left\{n_i^{\left(B\right)}\right\}$ at receiver
$B$ is assumed to be i.i.d complex Gaussian random variables with
$E{\left[\left|n_i^{\left(B\right)}\right|^2\right]} = N_0$, and
$\left|h_{A,B}\right|^2$ is exponentially distributed with mean
$E{\left[\left|h_{A,B}\right|^2\right]}$. Without loss of
generality, we assume that
$E{\left[\left|h_{A,B}\right|^2\right]}=1$. The SINR $C_{A,B}$ from
$A$ to $B$ is then given by

$$C_{A,B}=\frac{E_s\left|h_{A,B}\right|^2}{\sum_{A_i \in \mathcal
{R}}E_s{\left|h_{A_i,B}\right|^2}+N_0/2}$$

For a legitimate node and an eavesdropper, we use two separate SINR
thresholds $\gamma_R$ and $\gamma_E$ to define the minimum SINR
required to recover the transmitted messages for legitimate node and
eavesdropper, respectively. Therefore, a system node (relay or
destination) is able to decode a packet if and only if its SINR is
greater than $\gamma_R$, while the transmitted message is secure if
and only if the SINR at each eavesdropper is less than $\gamma_E$.

\subsection{Transmission Protocols}

We consider here two transmission protocols for secure and reliable
information transmission in two-hop wireless networks. The first
protocol (hereafter called Protocol 1) is the one proposed in
\cite{IEEEhowto:Goeckel}, in which the optimal relay node with the
best link condition to both source and destination is always
selected for information relaying. Although this protocol is
attractive in the sense that it provide very effective resistance
against eavesdroppers, it suffers from several problems. The
protocol 1 involves a complicated process of optimal relay
selection, which is not very suitable for the distributed wireless
networks, in particular when the number of possible relay nodes is
huge. More importantly, since the channel state is relatively
constant during a fixed time period, some relay nodes with good link
conditions are always preferred for information relaying, resulting
a severe load balance problem and a quick node energy depletion in
energy-limited wireless environment.

%However, these scheme either limit to a small network(source, relay, destination and eavesdroppers) or need to know eavesdropper channels and locations. By considering cooperative jamming with multiple relays,  But, this protocol has two disadvantages: unbalanced energy consumption among relays, which make some relay premature death because of energy depletion; the complex relay selection, which become even more daunting as the assistant nodes grows.ably as relays and energy consumption is not balanced among the candidate relays, which is unaccepted , such as wireless sensor networks. with the best links to both source and destination is always selected as relay, which makes transmission more effective and can resist more eavesdroppers. However, because the  Moreover, it is a complex work to decide the desired relay in the distributed wireless networks, which become even more daunting as the assistant nodes grows. In Protocol 1 is optimal to transmit messages, since the node with the best links to both source and destination is selected as relay. However, Protocol 1 presents challenges: unbalanced energy consumption and the complexity of the relay selection.

Based on these observations, we propose to use a simple and random
relay selection rather than the optimal relay selection to achieve a
better load and energy consumption balance among possible relay
nodes. By modifying the Protocol 1 to include the random relay
selection process, the new transmission protocol (hereafter called
Protocol 2) works as follows.

\textbf{1) \emph{Relay selection}:} A relay node, indexed by
$j^\ast$, is selected randomly from candidate relay nodes $R_j,
j=1,2,\cdots,n$.

\textbf{2) \emph{Channel measurement between the selected relay and
the other relays}:} The selected relay $j^\ast$ broadcasts a pilot
signal to allow each of other relays to measure the channel from
$j^\ast$ to itself. Each of the other relays $R_j, j=1,2,\cdots,n, j
\neq j^\ast$ then knows the corresponding value of
$h_{R_j,R_{j^\ast}}$.

\textbf{3) \emph{Channel measurement between destination $D$ and the
other relays}:} The destination $D$ broadcasts a pilot signal to
allow each of other relays to measure the channel from $D$ to
itself. Each of the other relays $R_j, j=1,2,\cdots,n, j \neq
j^\ast$ then knows the corresponding value of $h_{R_j,D}$.

\textbf{4) \emph{Message transmission from source $S$ to the
selected relay $R_{j^\ast}$}:} The source $S$ transmits the messages
to $R_{j^\ast}$. Concurrently, the relay nodes with indexes in
$\mathcal {R}_1 = {\left\{j \neq j^\ast : |h_{R_j,R_{j^\ast}}|^2 <
\tau \right\}}$, transmit noise to generate sufficient interference
at eavesdroppers.

\textbf{5) \emph{Message transmission from the selected relay
$R_{j^\ast}$ to destination $D$}:} Similar to the Step 4, the relay
$R_{j^\ast}$ transmits the messages to destination $D$.
Concurrently, the relay nodes with indexes in $\mathcal {R}_2 =
{\left\{j \neq j^\ast : |h_{R_j,D}|^2 < \tau \right\}}$, transmit
noise to generate sufficient interference at eavesdroppers.

\section{Theoretical Analysis}

This section first defines the transmission outage and secrecy
outage adopted in this paper to depict transmission reliability and
transmission secrecy, and then provides theoretical analysis to
determine the numbers of eavesdroppers a network can tolerate based
on the Protocol 1 and Protocol 2, respectively.

\subsection{Transmission Outage and Secrecy Outage}

For a transmission from the source $S$ to destination $D$, we call
transmission outage happens if $D$ can not decode the transmitted
packet, i.e., $D$ received the packet with SINR less than the
predefined threshold $\gamma_R$. The transmission outage
probability, denoted as $P_{out}^{\left(T\right)}$, is then defined
as the probability that transmission outage from $S$ to $D$ happens.
For a predefined upper bound $\varepsilon_t$ on
$P_{out}^{\left(T\right)}$, we call the communication between $S$
and $D$ is reliable if $P_{out}^{\left(T\right)} \leq
\varepsilon_t$. Notice that for the transmissions from $S$ to the
selected relay $R_{j^\ast}$ and from $R_{j^\ast}$ to $D$, the
corresponding transmission outage can be defined in the similar way
as that of from $S$ to $D$. We use $O_{S \rightarrow
R_{j^\ast}}^{(T)}$ and $O_{R_{j^\ast} \rightarrow D}^{(T)}$ to
denote the events that transmission outage from source $S$ to
$R_{j^\ast}$ happens and transmission outage from relay $R_{j^\ast}$
to $D$ happens, respectively. Due to the link independence
assumption, we have

\begin{align*}
&P_{out}^{\left(T\right)} =P\left(O_{S \rightarrow R_{j^\ast}}^{(T)}
\cup O_{R_{j^\ast} \rightarrow D}^{(T)}\right)\\
&\ \ \ \ \ \ = P\left(O_{S \rightarrow
R_{j^\ast}}^{(T)}\right)+P\left(O_{R_{j^\ast}
\rightarrow D}^{(T)}\right)\\
&\ \ \ \ \ \ \ \ \ -P\left(O_{S \rightarrow R_{j^\ast}}^{(T)}\right)
\cdot P\left(O_{R_{j^\ast} \rightarrow D}^{(T)}\right)
\end{align*}

%Therefore, $O_{S\rightarrow D} = \left(O_{S \rightarrow R_{j^\ast}} \cup O_{R_{j^\ast} \rightarrow D}\right)$. Define the transmission outage probability $P_{out}^{\left(T\right)}$ as the probability that the event $O_{S \rightarrow D}$ happens. Define the link transmission outage probability $P_{out}^{\left(S \rightarrow R_{j^\ast}\right)}$ as the probability that event $O_{S \rightarrow R_{j^\ast}}$ happens, and the link transmission outage probability $P_{out}^{\left(R_{j^\ast} \rightarrow D\right)}$ as the probability that event $O_{R_{j^\ast} \rightarrow D}$ happens. Because the all links are independent, $P_{out}^{\left(T\right)} = P_{out}^{\left(S \rightarrow R_{j^\ast}\right)} + P_{out}^{\left(R_{j^\ast}\rightarrow D\right)} - P_{out}^{\left(S \rightarrow R_{j^\ast}\right)} \cdot P_{out}^{\left(R_{j^\ast} \rightarrow D\right)}$.

Regarding the secrecy outage, we call secrecy outage happens for a
transmission from $S$ to $D$ if at least one eavesdropper can
recover the transmitted packets during the process of this two-hop
transmission, i.e., at least one eavesdropper received the packet
with SINR larger than the predefined threshold $\gamma_E$. The
secrecy outage probability, denoted as $P_{out}^{\left(S\right)}$,
is then defined as the probability that secrecy outage happens
during the transmission from $S$ to $D$. For a predefined upper
bound $\varepsilon_s$ on $P_{out}^{\left(S\right)}$, we call the
communication between $S$ and $D$ is secure if
$P_{out}^{\left(S\right)} \leq \varepsilon_s$. Notice that for the
transmissions from $S$ to the selected relay $R_{j^\ast}$ and from
$R_{j^\ast}$ to $D$, the corresponding secrecy outage can be defined
in the similar way as that of from $S$ to $D$. We use $O_{S
\rightarrow R_{j^\ast}}^{(S)}$ and $O_{R_{j^\ast} \rightarrow
D}^{(S)}$ to denote the events that secrecy outage from source $S$
to $R_{j^\ast}$ happens and secrecy outage from relay $R_{j^\ast}$
to $D$ happens, respectively. Again, due to the link independence
assumption, we have

\begin{align*}
&P_{out}^{\left(S\right)} =P\left(O_{S \rightarrow
R_{j^\ast}}^{(S)}\right)+P\left(O_{R_{j^\ast} \rightarrow
D}^{(S)}\right)\\
&\ \ \ \ \ \ \ \ \ -P\left(O_{S \rightarrow R_{j^\ast}}^{(S)}\right)
\cdot P\left(O_{R_{j^\ast} \rightarrow D}^{(S)}\right)
\end{align*}

\subsection{Analysis of Protocol 1}
In the Protocol 1 proposed in \cite{IEEEhowto:Goeckel}, the relay
node with the largest value of
$min\left(\left|h_{S,R_j}\right|^2,\left|h_{D,R_j}\right|^2\right),
j= 1, 2, \cdots, n$, is selected as relay. Notice that the Protocol
1 can always guarantee the reliable transmission from source $S$ to
destination $D$, this is because the parameter $\tau$ is set as
$\tau = \sqrt{\frac{\log{n}}{8n\gamma_R}}$, which ensures that the
target SINR at the selected relay and destination can be achieved to
decode the transmitted messages. Thus, we only need to focus the
secrecy requirement $P_{out}^{\left(S\right)} \leq \varepsilon_s$ to
determine the corresponding the number of eavesdroppers the network
can tolerate here.

\textbf{Theorem 1.} For the network scenario illustrated in Fig 1
with equal path loss between all pairs of nodes, to guarantee the
secrecy requirement $P_{out}^{(S)} \leq \varepsilon_s$ by applying
the Protocol 1, the number of eavesdroppers $m$ the network can
tolerate should satisfy the following condition.
$$m \leq \left(1- \sqrt{1-\varepsilon_s}\right) \cdot
\left(1+\gamma_E\right)^{\sqrt{\frac{n\log{n}}{32\gamma_R}}}$$

\begin{proof}

Notice that $P_{out}^{\left(S\right)}$ is determined as
\begin{align*}
&P_{out}^{\left(S\right)} = P\left(O_{S \rightarrow
R_{j^\ast}}^{(S)}\right)+P\left(O_{R_{j^\ast} \rightarrow
D}^{(S)}\right)\\
&\ \ \ \ \ \ \ \ \ -P\left(O_{S \rightarrow R_{j^\ast}}^{(S)}\right)
\cdot P\left(O_{R_{j^\ast} \rightarrow D}^{(S)}\right)
\end{align*}

Since the transmission process from source $S$ to the selected relay
$R_{j^\ast}$ is identical to that of from the selected relay
$R_{j^\ast}$ to destination $D$, we have

$$P\left(O_{S \rightarrow
R_{j^\ast}}^{(S)}\right) = P\left(O_{R_{j^\ast} \rightarrow
D}^{(S)}\right)$$ and
$$P_{out}^{\left(S\right)} = 2P\left(O_{S \rightarrow
R_{j^\ast}}^{(S)}\right)
 - \left[P\left(O_{S \rightarrow
R_{j^\ast}}^{(S)}\right)\right]^2 $$

To ensure $P_{out}^{\left(S\right)} \leq \varepsilon_s$, then should
have

$$P\left(O_{S \rightarrow R_{j^\ast}}^{(S)}\right) \leq 1- \sqrt{1-\varepsilon_s}$$

From the reference \cite{IEEEhowto:Goeckel}, we notice that

\begin{align*}
& P\left(O_{S \rightarrow R_{j^\ast}}^{(S)}\right) =
P\left(\bigcup_{i=1}^{m}\left\{C_{S,E_i} \geq
\gamma_E\right\}\right)\\
& \ \ \ \ \ \ \ \ \ \ \ \ \ \ \ \ \ \leq
\sum_{i=1}^{m}P\left(C_{S,E_i} \geq
\gamma_E\right)\\
& \ \ \ \ \ \ \ \ \ \ \ \ \ \ \ \ \ \leq
m\cdot\left(\frac{1}{1+\gamma_E}\right)^{\sqrt{\frac{n\log{n}}{32\gamma_R}}}
\end{align*}

To guarantee the secrecy requirement, we just need
\begin{align*}
&m\cdot\left(\frac{1}{1+\gamma_E}\right)^{\sqrt{\frac{n\log{n}}{32\gamma_R}}}
\leq 1- \sqrt{1-\varepsilon_s}
\end{align*}

and thus
\begin{align*}
& m \leq \left(1- \sqrt{1-\varepsilon_s}\right) \cdot
\left(1+\gamma_E\right)^{\sqrt{\frac{n\log{n}}{32\gamma_R}}}
\end{align*}

\end{proof}

\subsection{Analysis of Protocol 2}

The parameter $\tau$ involved in the Protocol 2 determines whether
the relay and destination can receive the messages successfully and
whether sufficient noise is generated to suppress eavesdroppers. For
the analysis of the Protocol 2, we first determine the range for the
parameter $\tau$ to ensure both secrecy requirement and reliability
requirement, based on which we then analyze the number of
eavesdroppers a network can be tolerate by applying the protocol.

\textbf{Theorem 2.} Consider the network scenario of Fig 1 with
equal path loss between all pairs of nodes, to ensure
$P_{out}^{\left(T\right)} \leq \varepsilon_t$ and
$P_{out}^{\left(S\right)} \leq \varepsilon_s$ by applying the
Protocol 2, the parameter $\tau$ must satisfy the following
condition.

$$\tau \in \left[- \log{\left[1 + \frac{\log{\left(\frac{1 - \sqrt{1 -
\varepsilon_s}}{m}\right)}}{\left(n - 1\right)\log{\left(1 +
\gamma_E\right)}}\right]},
\sqrt{\frac{-\log\left(1-\varepsilon_t\right)}{2\gamma_R \left(n
-1\right)}}\right]$$

\begin{proof}

Notice that in the Protocol 2, a larger value of $\tau$ indicates
that more system nodes will generate noise to suppress the
eavesdroppers. However, too high noise will also interrupt the
legitimate transmission. Therefore, the parameter $\tau$ should be
set properly to satisfy both reliability and secrecy requirements.

\textbf{$\bullet$ Reliability Guarantee}

Notice that $P_{out}^{\left(T\right)}$ is determined as
\begin{align*}
&P_{out}^{\left(T\right)} = P\left(O_{S \rightarrow
R_{j^\ast}}^{(T)}\right)+P\left(O_{R_{j^\ast} \rightarrow
D}^{(T)}\right)\\
&\ \ \ \ \ \ \ \ \ \ -P\left(O_{S \rightarrow
R_{j^\ast}}^{(T)}\right) \cdot P\left(O_{R_{j^\ast} \rightarrow
D}^{(T)}\right)
\end{align*}

Because the transmission process from source $S$ to the selected
relay $R_{j^\ast}$ is identical to that of from the selected relay
$R_{j^\ast}$ to destination $D$, we have

$$P\left(O_{S
\rightarrow R_{j^\ast}}^{(T)}\right) = P\left(O_{R_{j^\ast}
\rightarrow D}^{(T)}\right)$$

and

$$P_{out}^{\left(T\right)}= 2P\left(O_{S
\rightarrow R_{j^\ast}}^{(T)}\right)
 - \left[P\left(O_{S
\rightarrow R_{j^\ast}}^{(T)}\right)\right]^2 $$

To ensure $P_{out}^{\left(T\right)} \leq \varepsilon_t$, we need

$$P\left(O_{S
\rightarrow R_{j^\ast}}^{(T)}\right) \leq 1-
\sqrt{1-\varepsilon_t}$$

Based on the definition of transmission outage probability, we have

\begin{align*}
& P\left(O_{S
\rightarrow R_{j^\ast}}^{(T)}\right)\\
& \ \ \ \ \ = P\left(C_{S,R_{j^\ast}} \leq
\gamma_R\right)\\
& \ \ \ \ \ = P\left(\frac{E_s \cdot |h_{S,R_{j^\ast}}|^2}{\sum_{R_j
\in \mathcal {R}_1}E_s \cdot
|h_{R_j,R_{j^\ast}}|^2 + N_0/2} \leq \gamma_R\right)\\
& \ \ \ \ \ \doteq P\left(\frac{|h_{S,R_{j^\ast}}|^2}{\sum_{R_j \in
\mathcal
{R}_1}|h_{R_j,R_{j^\ast}}|^2} \leq \gamma_R\right)\\
\end{align*}

Compared to the noise generated by multiple system nodes, the
environment noise is negligible and thus is omitted here to simply
the analysis. Notice that $\mathcal {R}_1 = {\left\{j \neq j^\ast :
|h_{R_j,R_{j^\ast}}|^2 < \tau \right\}}$, then

\begin{align*}
& P\left(O_{S \rightarrow R_{j^\ast}}^{(T)}\right) \leq
P\left(\frac{|h_{S,R_{j^\ast}}|^2}{{|\mathcal {R}_1|}\tau} \leq
\gamma_R\right)\\
& \ \ \ \ \ \ \ \ \ \ \ \ \ \ \ \ \ = P\left(|h_{S,R_{j^\ast}}|^2 \leq \gamma_R{|\mathcal {R}_1|}\tau\right)\\
& \ \ \ \ \ \ \ \ \ \ \ \ \ \ \ \ \ = 1 - e^{-\gamma_R{|\mathcal
{R}_1|}\tau}
\end{align*}

Since there are $n - 1$ other relays except $R_{j^\ast}$, the
expected number of noise-generation nodes is given by $|\mathcal
{R}_1| =\left(n-1\right) \cdot P\left(|h_{R_j,R_{j^\ast}}|^2 <
\tau\right) = \left(n-1\right) \cdot \left(1-e^{-\tau}\right)$. Then
we have
\begin{align*}
&P\left(O_{S \rightarrow R_{j^\ast}}^{(T)}\right) \leq 1 -
e^{-\gamma_R\left(n -1\right)\left(1-e^{-\tau}\right)\tau}
\end{align*}

Thus, to ensure reliability requirement, we just need

$$1 - e^{-\gamma_R\left(n -1\right)\left(1-e^{-\tau}\right)\tau} \leq 1 - \sqrt{1 -
\varepsilon_t}$$

That is,

$$-\gamma_R \left(n -1\right) \left(1-e^{-\tau}\right) \tau \geq \frac{1}{2}\log\left(1-\varepsilon_t\right)$$

By using Taylor formula, we have

$$\tau^2 \leq \frac{-\log\left(1-\varepsilon_t\right)}{2\gamma_R \left(n -1\right)} $$

and thus

\begin{align*}
\tau \leq \sqrt{\frac{-\log\left(1-\varepsilon_t\right)}{2\gamma_R
\left(n -1\right)}}
\end{align*}

The above result indicates that
$\sqrt{\frac{-\log\left(1-\varepsilon_t\right)}{2\gamma_R \left(n
-1\right)}}$ is the maximum value the parameter $\tau$ can take to
ensure the reliability requirement.

\textbf{$\bullet$ Secrecy Guarantee}

Notice that $P_{out}^{\left(S\right)}$ is given by

\begin{align*}
&P_{out}^{\left(S\right)} = P\left(O_{S \rightarrow
R_{j^\ast}}^{(S)}\right)+P\left(O_{R_{j^\ast} \rightarrow
D}^{(S)}\right)\\
&\ \ \ \ \ \ \ \ \ -P\left(O_{S \rightarrow R_{j^\ast}}^{(S)}\right)
\cdot P\left(O_{R_{j^\ast} \rightarrow D}^{(S)}\right)
\end{align*}

Since the transmission process from source $S$ to the selected relay
$R_{j^\ast}$ is identical to that of from the selected relay
$R_{j^\ast}$ to destination $D$, then we have

$$P\left(O_{S \rightarrow
R_{j^\ast}}^{(S)}\right) = P\left(O_{R_{j^\ast} \rightarrow
D}^{(S)}\right)$$

and

$$P_{out}^{\left(S\right)} = 2P\left(O_{S \rightarrow
R_{j^\ast}}^{(S)}\right)
 - \left[P\left(O_{S \rightarrow
R_{j^\ast}}^{(S)}\right)\right]^2 $$

To ensure $P_{out}^{\left(S\right)} \leq \varepsilon_s$, we need

$$P\left(O_{S \rightarrow R_{j^\ast}}^{(S)}\right) \leq 1- \sqrt{1-\varepsilon_s}$$

According to the definition of secrecy outage probability, we know
that

\begin{align*}
&P\left(O_{S \rightarrow R_{j^\ast}}^{(S)}\right) =
P\left(\bigcup_{i=1}^{m}\left\{C_{S,E_i} \geq
\gamma_E\right\}\right)
\end{align*}

Thus, we have

\begin{align*}
&P\left(O_{S \rightarrow R_{j^\ast}}^{(S)}\right) \leq
\sum_{i=1}^{m}P\left(C_{S,E_i} \geq \gamma_E\right)
\end{align*}

Based on the Markov inequality,

\begin{align*}
& P\left(C_{S,E_i} \geq \gamma_E\right)\\
& \ \ \ \ \ \leq P\left(\frac{E_s \cdot |h_{S,E_i}|^2}{\sum_{R_j \in
\mathcal {R}_1}E_s \cdot |h_{R_j,E_i}|^2} \geq
\gamma_E\right)\\
& \ \ \ \ \ = E_{\left\{h_{R_j,E_i}, j=0,1,\cdots,n+mp,j \neq
j^{\ast}\right\},\mathcal {R}_1}\\
& \ \ \ \ \ \ \ \ \ \left[P\left(|h_{S,E_i}|^2 > \gamma_E
\cdot \sum_{R_j \in \mathcal {R}_1}|h_{R_j,E_i}|^2\right)\right]\\
& \ \ \ \ \ \leq E_{\mathcal {R}_1}\left[\prod_{R_j \in \mathcal
{R}_1}
E_{h_{R_j,E_i}}\left[e^{-\gamma_E|h_{R_j,E_i}|^2}\right]\right]\\
& \ \ \ \ \ = E_{\mathcal
{R}_1}\left[\left(\frac{1}{1+\gamma_E}\right)^{|\mathcal
{R}_1|}\right]
\end{align*}

Therefore,

$$P\left(O_{S \rightarrow R_{j^\ast}}^{(S)}\right) \leq
\sum_{i=1}^{m}\left(\frac{1}{1+\gamma_E}\right)^{|\mathcal {R}_1|} =
m \cdot \left(\frac{1}{1+\gamma_E}\right)^{|\mathcal {R}_1|}$$

To ensure the secrecy requirement, we just need

\begin{align*}
&m \cdot \left(\frac{1}{1+\gamma_E}\right)^{|\mathcal {R}_1|} \leq
1- \sqrt{1-\varepsilon_s}
\end{align*}

or equally

\begin{align*}
& \left(\frac{1}{1+\gamma_E}\right)^{\left(n
-1\right)\left(1-e^{-\tau}\right)} \leq \frac{1 - \sqrt{1 -
\varepsilon_s}}{m}
\end{align*}

\begin{align*}
&\left(n -1\right)\left(1-e^{-\tau}\right) \geq -
\frac{\log{\left(\frac{1 - \sqrt{1 -
\varepsilon_s}}{m}\right)}}{\log{\left(1 + \gamma_E\right)}}
\end{align*}

\begin{align*}
& e^{-\tau} \leq 1 + \frac{\log{\left(\frac{1 - \sqrt{1 -
\varepsilon_s}}{m}\right)}}{\left(n - 1\right)\log{\left(1 +
\gamma_E\right)}}
\end{align*}

Then we have

\begin{align*}
& \tau \geq - \log{\left[1 + \frac{\log{\left(\frac{1 - \sqrt{1 -
\varepsilon_s}}{m}\right)}}{\left(n - 1\right)\log{\left(1 +
\gamma_E\right)}}\right]}
\end{align*}

The above result shows that $ - \log{\left[1 +
\frac{\log{\left(\frac{1 - \sqrt{1 -
\varepsilon_s}}{m}\right)}}{\left(n - 1\right)\log{\left(1 +
\gamma_E\right)}}\right]}$ is the minimum value parameter $\tau$ can
take to guarantee the secrecy requirement.

\end{proof}

Based on the results of Theorem 2, we now can establish the
following theorem about the performance of Protocol 2.

\textbf{Theorem 3.} Consider the network scenario of Fig 1 with
equal path loss between all pairs of nodes. To guarantee
$P_{out}^{\left(T\right)} \leq \varepsilon_t$ and
$P_{out}^{\left(S\right)} \leq \varepsilon_s$ based on the Protocol
2, the number of eavesdroppers $m$ the network can tolerate must
satisfy the following condition.

\begin{align*}
& m \leq \left(1 - \sqrt{1 -
\varepsilon_s}\right)\cdot\left(1+\gamma_E\right)^{\sqrt{\frac{-\left(n-1\right)\log\left(1-\varepsilon_t\right)}{2\gamma_R}}}
\end{align*}

\begin{proof}

%From the Theorem 2, when $\tau =\sqrt{\frac{-\log\left(\frac{\sqrt{1-\varepsilon_t}}{2}\right)}{\gamma_R \left(n -1\right)}}$, the transmission is exactly reliable. The larger $\tau$ means more assistant nodes are selected to generate noise to suppress the eavesdroppers. When $\tau$ gets the maximum value, the number of eavesdroppers which can be tolerated is maximum.

From Theorem 2 we know that to ensure the reliability requirement,
we have

\begin{align*}
\tau \leq \sqrt{\frac{-\log\left(1-\varepsilon_t\right)}{2\gamma_R
\left(n -1\right)}}
\end{align*}

and

\begin{align*}
& \left(n - 1\right)\left(1 - e^{-\tau}\right) \leq
\frac{-\log\left(1-\varepsilon_t\right)}{2\gamma_R \tau}
\end{align*}

To ensure the secrecy requirement, we need

\begin{align*}
& \left(\frac{1}{1+\gamma_E}\right)^{\left(n
-1\right)\left(1-e^{-\tau}\right)} \leq \frac{1 - \sqrt{1 -
\varepsilon_s}}{m}
\end{align*}

Thus,

\begin{align*}
& m \leq \frac{1 - \sqrt{1 -
\varepsilon_s}}{\left(\frac{1}{1+\gamma_E}\right)^{\left(n
-1\right)\left(1-e^{-\tau}\right)}}\\
&\ \ \ \ \leq \frac{1 - \sqrt{1 -
\varepsilon_s}}{\left(\frac{1}{1+\gamma_E}\right)^{\frac{-\log\left(1-\varepsilon_t\right)}{2\gamma_R
\tau}}}
\end{align*}

By let $\tau$ taking its maximum value, we get the following bound

\begin{align*}
& m \leq \frac{1 - \sqrt{1 -
\varepsilon_s}}{\left(\frac{1}{1+\gamma_E}\right)^{\sqrt{\frac{-\left(n-1\right)\log\left(1-\varepsilon_t\right)}{2\gamma_R}}}}
\end{align*}

That is,

\begin{align*}
& m \leq \left(1 - \sqrt{1 -
\varepsilon_s}\right)\cdot\left(1+\gamma_E\right)^{\sqrt{\frac{-\left(n-1\right)\log\left(1-\varepsilon_t\right)}{2\gamma_R}}}
\end{align*}

\end{proof}

\subsection{Discussion}
The two protocols considered in this paper have their own advantages
and disadvantages and thus are suitable for different network
scenarios. For the protocol 1 proposed in \cite{IEEEhowto:Goeckel},
it can achieve a better performance in terms of the number of
eavesdroppers can be tolerated. However, such protocol always tend
to select the optimal node with the best links to both source and
destination as the relay, so it involves a complex relay selection
process, and more importantly, it results in an unbalanced load and
energy consumption distribution among systems nodes. Thus, such
protocol is suitable for small scale wireless network with
sufficient energy supply rather than large and energy-limited
wireless networks (like wireless sensor networks).
 Regarding the Protocol 2, although it can tolerate less number eavesdroppers in comparison with the Protocol 1, it involves a very simple random relay selection process to achieve a very good load and energy consumption distribution among system nodes. Thus, this protocol is more suitable for large
scale wireless network environment with stringent energy consumption
constraint.

%In \cite{IEEEhowto:Goeckel}, though the cooperative relay protocol is analyzed as the number of assistant nodes tends to infinity, the optimal relay selection is very complex and difficult in the large scale wireless networks. For energy-limited wireless environment, such as wireless sensor networks, some relays nodes with the optimal links to both the source and destination will die prematurely due to energy depletion, since such assistant nodes are always selected as relays to transmit messages. Node premature death is not tolerated in cooperative relay wireless networks, which will lead to no enough cooperative nodes.

%and it randomly selects a assistant node as relay randomly, which can balance energy consumption among the candidate relays and simplify the relay selection process while achieving the desired secrecy. Since the relay selection does not become more complex as the network size grows, protocol 2 is more suitable for the large scale wireless network environment. Moreover, protocol 2 can be used in the energy-limited wireless sensor networks, because it balances the energy consumption among assistant nodes, and avoids some nodes premature death due to energy depletion. These two protocols have their own advantages and disadvantages. It is the more effectively information transmission scheme with the joint use of these two protocols according to the application scenario, since protocol 1 is suitable for the small size and sufficient energy wireless networks and protocol 2 is more suitable for the large size and limited energy wireless networks.

\section{Conclusion}

This paper explores reliable and secure information transmission
through multiple cooperative systems nodes in two-hop relay wireless
network with passive eavesdroppers of unknown channels and
locations, for which two transmission protocols are considered. For
each protocol, theoretical analysis has been provided to show the
number of eavesdroppers the network can tolerate subject to
constraints on transmission outage probability and secrecy outage
probability. These two protocols, each has different performance in
terms of eavesdropper tolerance, load and energy consumption
distribution among nodes, and also relay selection complexity, are
suitable for different network scenarios depending on network scale
and also energy consumption constraint there.

%It is the more effectively information transmission scheme with the joint use of these two protocols according to the application scenario, since protocol 1 is suitable for the small size and sufficient energy wireless networks and protocol 2 is more suitable for the large size and limited energy wireless networks  By employing artificial noise generation and random relay selection techniques, the protocol proposed in \cite{IEEEhowto:Goeckel} is improved to ensure the information transmission secrecy, balance the energy consumption among the candidate relays and simplify the relay selection process in two-hop relay wireless networks. We analyze the advantages and disadvantages of these two protocols proposed in \cite{IEEEhowto:Goeckel} and this paper respectively, determine the precise number of passive eavesdroppers which can be tolerated for each protocol and discuss their application scenarios. This result can guide the design and deployment of the actual wireless networks. In the further works, We will consider the scene in presence of both passive and active eavesdroppers.

%\section*{Acknowledgment}
%Supported by the National Science Foundation of China(61100153, 61172068, U0835004) and the Fundamental Research Funds for the Central Universities (K5051203008).


\ifCLASSOPTIONcaptionsoff
\newpage
\fi



%\bibliographystyle{IEEEtran}

\begin{thebibliography}{14}

\bibitem{IEEEhowto:Sathya}
Narayanan, Sathya, Two-hop forwarding in wireless networks,
dissertation for the degree of Doctor of philosophy, Polytechnic
University, 2006

\bibitem{IEEEhowto:Talbot}
J. Talbot and D. Welsh, \emph{Complexity and Crytography : An
Introduction}, Cambridge, 2006.

\bibitem{IEEEhowto:Morr}
K. Morrison, and D. Goeckel£¬\emph{Power allocation to
noise-generating nodes for cooperative secrecy in the wireless
environment}. In the Forty Fifth Asilomar Conference on Signals,
Systems and Computers (ASILOMAR), 275-279, 2011.

\bibitem{IEEEhowto:Goel}
S. Goel, and R. Negi, \emph{Guaranteeing secrecy using artificial
noise}. IEEE transactions on wireless communications,
7(6):2180-2189, 2008.

\bibitem{IEEEhowto:Lai}
L. Lai and H. El Gamal, \emph{The relay-eavesdropper channel:
Cooperation for secrecy}, IEEE Trans. Inf. Theory, vol. 54, no. 9,
pp. 4005 - 4019, Sept. 2008.

\bibitem{IEEEhowto:Yuksel}
M. Yuksel and E. Erkip, \emph{Secure communication with a relay
helping the wiretapper}, in Proc. 2007 IEEE Information Theory
Workshop, Lake Tahoe, CA, Sept. 2007.

\bibitem{IEEEhowto:Negi}
R. Negi and S. Goelm, \emph{Secret communication using artificial
noise}, in Proc. IEEE Vehicular Tech. Conf, vol. 3, Dallas TX, pp.
1906-1910, Sept. 2005.


\bibitem{IEEEhowto:Vasudevan}
S. Vasudevan, S. Adams, D. Geockel, Z. Ding, D. Towsley, and K.
Leung, \emph{Multi-user diversity for secrecy in wireless networks}.
In Information Theorem and Applications Workshop, 2009.

\bibitem{IEEEhowto:He}
X. He and A. Yener, \emph{Two-hop secure communication using an
untrusted relay: A case for cooperative jamming}, in Proc. 2008 IEEE
Global Telecommunications Conference, New Orleans, LA, Nov. - Dec.
2008.

\bibitem{IEEEhowto:Dong}
L. Dong, Z. Han, A. Petropulu, and H. V. Poor, \emph{Improving
wireless physical layer security via cooperating relays}, IEEE
Trans. Sig. Proc., vol. 58, no. 3, pp. 1875-1888, Mar. 2010.

\bibitem{IEEEhowto:Goeckel}
D. Goeckel, S. Vasudevan, D. Towsley, S. Adams, Z. Ding, and K.
Leung, \emph{Artificial noise generation from cooperative relays for
everlasting secrecy in two-hop wireless networks,¡± IEEE Journal on
Selected Areas in Communications}, 29(10):2067-2076, 2011.

\end{thebibliography}

\end{document}